\documentclass[final]{aipproc}
\layoutstyle{6x9}%
\usepackage{amsfonts}
\usepackage{amsmath}
\usepackage[applemac]{inputenc}

\begin{document}

\title{On the role of the chaotic velocity in relativistic kinetic theory}


\author{Valdemar Moratto}{
  address={Departamento de F\'isica, Universidad Aut\'onoma Metropolitana-Iztapalapa, M\'exico D.F., M\'exico.}
}

\author{A. L. Garc\'ia-Perciante}{
  address={Departamento de Matem\'aticas Aplicadas y Sistemas, Universidad Aut\'onoma Metropolitana-Cuajimalpa, M\'exico D.F., M\'exico.}
}

\begin{abstract}
In this paper we revisit the concept of chaotic velocity within the
context of relativistic kinetic theory. Its importance as the key
ingredient which allows to clearly distinguish convective and dissipative
effects is discussed to some detail. Also, by addressing the case
of the two component mixture, the relevance of the barycentric comoving
frame is established and thus the convenience for the introduction of peculiar
velocities for each species. The fact that the decomposition of molecular
velocity in systematic and peculiar components does not alter the
covariance of the theory is emphasized. Moreover, we show that within
an equivalent decomposition into space-like and time-like tensors,
based on a generalization of the relative velocity concept, the Lorentz
factor for the chaotic velocity can be expressed explicitly as an
invariant quantity. This idea, based on Ellis' theorem, allows to
foresee a natural generalization to the general relativistic case.
\end{abstract}

\maketitle

\section{Introduction}

The concept of chaotic velocity was introduced in non-relativistic
kinetic theory since its early developments \cite{Brush,Maxwell}.
Its importance in the formulation of the theory, as well as its connection
with the corresponding phenomenology, resides in the fact that it
allows to separate mechanic and thermodynamic effects. Moreover, the
heat flux was defined by Clausius \cite{Brush} as the flow of energy
arising from the purely chaotic component of the motion. Eventhough
the chaotic velocity is a standard tool in the non relativistic formalism,
it has been mostly ignored in the relativistic case. The first work
recognizing its value and the need to include it in the relativistic
formulation in order to clearly define dissipative fluxes was written
by Sandoval and Garc\'ia-Col\'in \cite{Sandoval physica A 2000}. In that
work, Lorentz transformations were introduced with the purpose of
extracting the peculiar component from the molecular velocity. Such
idea was carried further in several publications, in particular Ref.
\cite{Garcia Perciante 2012} formulates a covariant kinetic theory
in terms of hydrodynamic and chaotic velocities in the framework of
special relativity. Also, Ref. \cite{Garcia-Perciante Chaotic Velocity}
includes a somehow thorough discussion and conceptual explanation
of such a decomposition.
\\
In this work we review the arguments presented in the references cited
above and provide with two additional contributions. First, we explicitly
show how the introduction of Lorentz transformations and the use of
the Lorentz factor $\gamma$ for one particle evaluated in the comoving
frame of the fluid as the key variable, preserve the covariance of
the theory. The use of such factor allows to express all variables
and fluxes in a similar way to the non-relativistic case. The particular
case of a binary mixture is addressed, in which the chaotic velocity
of the species plays a critical role as well as the system moving
with the barycentric velocity. Secondly, we show how by introducing
the concept of relative velocity, proposed in a different context
by Ellis \cite{ellis articulo,ellis libro}, one is able to split
the hydrodynamic velocity into two components in such a way that the
space-like part permits the expression of the $\gamma$ factor in
a covariant way for a general metric.
\\
The structure of this paper is as follows. In the second section we
make a brief review of what is the meaning and implications of the
use of the chaotic velocity in non-relativistic kinetic theory. The
third section addresses the introduction of the chaotic velocity in
special relativity and discusses the case of the binary mixture. The
decomposition in terms of a space-like relative velocity is introduced
in the fourth section as a suitable alternative for extending the
concept to the framework of general relativity. The last section includes
concluding remarks and perspectives.
\\

\section{The chaotic velocity in non-relativistic kinetic theory}

As mentioned above, the importance of chaotic velocity, also known
as peculiar or thermal velocity, in non relativistic kinetic theory
is due to the fact that it allows one to separate diffusive and convective
effects. The relevance of this property can be appreciated by considering
for example the heat flux, which is the energy flux due to the molecular
nature of matter. By introducing such concept when analyzing energy
transport, mechanical contributions arising from the motion of the
system as a whole become separated from those related to thermal agitation.

To illustrate the point we review the discussion in Ref. \cite{Garcia Perciante 2012} and \cite{Chapman} by starting with the non relativistic Boltzmann equation for a simple gas in the absence of external forces, this is
\begin{equation}
\frac{\partial f}{\partial t}+\vec{v}\cdot\nabla f=J\left(ff'\right).\label{eq:2.1}
\end{equation}
Here $f=f\left(\vec{r},\vec{v},t\right)$ is the distribution function
per particle, $\vec{v}$ is the velocity of one particle with mass
$m$, as measured by an observer in the laboratory frame. The term
on the right hand side accounts for the variations in the distribution
function due to particle collisions. From Eq. (\ref{eq:2.1}) one
can obtain the balance equations by using the standard method, that
is, by multiplying it by the collisional invariants, namely the mass,
momentum and energy, and then integrating over velocity space \cite{Chapman}.
This process leads to the definitions, with the help of the local
equilibrium assumption, of the thermodynamic local variables as well
as the corresponding fluxes as averages over the distribution function.
For these definitions to be in accordance with the phenomenology and
the physical interpretation of such quantities, it is crucial to introduce
the decomposition
\begin{equation}
\vec{v}=\vec{c}+\vec{u},\label{eq:2.6-1}
\end{equation}
where $\vec{u}$ and $\vec{c}$ are the hydrodynamic and chaotic velocities
respectively. The procedure is standard and leads to the following
definitions for the state variables particle density, hydrodynamic
velocity and internal energy
\begin{equation}
n\left(\vec{r},t\right)=\int fd\vec{v},\label{eq:2.3}
\end{equation}
\begin{equation}
\vec{u}\left(\vec{r},t\right)=\frac{1}{n}\int\vec{v}fd\vec{v},\label{eq:2.4}
\end{equation}
and
\begin{equation}
ne\left(\vec{r},t\right)=\frac{1}{2}m\int c^{2}fd\vec{c},\label{eq:2.8.1}
\end{equation}
respectively. Notice how the internal energy arises solely from the
chaotic component of the velocity. The total energy is given by
\begin{equation}
\frac{1}{2}m\int v^{2}fd\vec{c}=\frac{1}{2}mnu^{2}+ne,
\end{equation}
where Eq. (\ref{eq:2.6-1}) has been introduced for $\vec{v}$ and
use has been made of the fact that, in view of Eq. (\ref{eq:2.4}),
$\int\vec{c}fd\vec{c}=0$. Regarding the fluxes one has, firstly for
the stress tensor
\begin{equation}
\overleftrightarrow{\tau}\left(\vec{r},t\right)=m\int\vec{v}\vec{v}fd\vec{c}
=m\int\vec{c}\vec{c}fd\vec{c}+\rho\vec{u}\vec{u},\label{eq:2.7}
\end{equation}
where $\rho=mn$. Note that the first term on the right hand side
of Eq. (\ref{eq:2.7}) refers strictly to the chaotic velocity $\vec{c}$
and not the molecular one $\vec{v}$. From this term, both the hydrostatic
pressure part as well as the one related with viscous dissipation
will arise. Due to the use of Eq. (\ref{eq:2.6-1}) the contribution
of the hydrodynamic velocity $\vec{u}$ to the stress tensor is driven
to the convective term $\rho\vec{u}\vec{u}$. Secondly, from the energy
balance equation, which is given by
\begin{equation}
\rho\frac{de}{dt}+\nabla\cdot\vec{J}_{q}+\left(\overleftrightarrow{\tau}^{k}\right):\nabla\vec{u}=0,\label{eq:2.8}
\end{equation}
one identifies the heat flux
\begin{equation}
\vec{J}_{q}\left(\vec{r},t\right)=\frac{1}{2}m\int c^{2}\vec{c}fd\vec{c},\label{eq:2.9}
\end{equation}
and the tensor associated with the viscosity
\begin{equation}
\overleftrightarrow{\tau}^{k}\left(\vec{r},t\right)=m\int\vec{c}\vec{c}fd\vec{v},\label{eq:2.10}
\end{equation}
which appear as dissipative fluxes in such an equation. We know from
thermostatics that the internal energy is defined with no contributions
from the motion of the system as a whole. Equation (\ref{eq:2.8})
is clearly consistent with this, since in it internal energy is dissipated
only by the heat flux $\vec{J}_{q}$ and the viscous contribution
$\overleftrightarrow{\tau}^{k}$ which only depend on the chaotic
velocity.

Having reviewed these definitions, it is clear that the decomposition
given by Eq. (\ref{eq:2.6-1}) is crucial for the physical interpretation
of the different contributions to fluxes and thermodynamic variables.
Notice that the separation of chaotic and hydrodynamic velocities
within the molecular one can also be interpreted as a change of reference
frame. This idea permits the extension to the framework of special
relativity which will be the focus of the next section. This interpretation
is discussed in Ref. \cite{Garcia-Perciante Chaotic Velocity} for
the simple fluid, we repeat the argument here in order to generalize
it to the fluid mixture.

For the simple fluid, the molecular velocity $\vec{v}$ is the velocity
of the molecule as measured in a laboratory frame while the hydrodynamic
velocity $\vec{u}$ is the field of velocities that defines the motion
of the fluid with respect to it. The chaotic velocity $\vec{c}$ is
thus clearly the velocity of the particles measured in a comoving
frame which is at rest with respect to the fluid element. In other
words, an observer in a locality on this comoving frame will observe
the fluid at rest in average and only measure the peculiar velocity
of the molecules. These ideas are illustrated in Fig. 1 where $\bar{S}$
is the comoving frame, in which an observer will measure as the molecular
velocity of some particle only the chaotic component $\vec{c}$. On
the other hand an observer in the $S$ frame will measure for the
same particle its corresponding Galilean transformed velocity, i.e.
$\vec{v}=\vec{c}+\vec{u}$.

\begin{figure}
\includegraphics[scale=0.5]{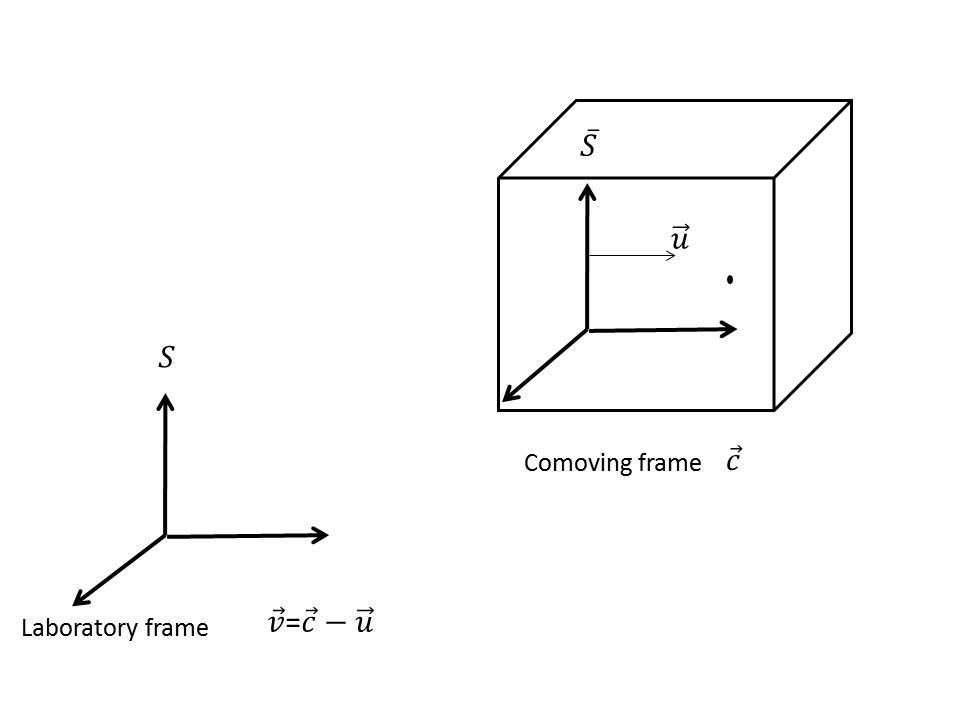}
\caption{
Comoving frame {$\bar{S}$} moving with velocity
$\vec{u}$ respect to the laboratory frame $S$. The chaotic velocity
is $\vec{c}$.
}
\end{figure}

An observer in the $S$ frame will obtain in his measurements not
only the dissipative effects driven from the molecular nature of the
fluid but also those convective contributions that arise form the
motion of the fluid as one big mechanical object. On the other hand,
one observer in the comoving frame will only measure the thermal effects.
This frame is convenient when we are interested in the evaluation
of thermodynamic properties because it leads in a very natural way
to the conceptualization of thermodynamics from the kinetic theory
point of view.

For the case of mixtures, the concept of chaotic velocity has another
feature which underlines its importance. In such a case, the definition
is given by
\begin{equation}
\vec{c}_{i}=\vec{v}_{i}-\vec{u}.\label{eq:2.11}
\end{equation}
Here the subindex $i$ denotes the velocity for the $i$-th species
in the gas ($i=\left\{ 1,...,M\right\} $) with $M$ constituents
and $\vec{u}\left(\vec{r},t\right)$ is the barycentric velocity of
the gas defined by
\begin{equation}
n\vec{u}=\sum_{i=1}^{M}n_{i}\vec{u}_{i},\label{eq:2.12}
\end{equation}
being $n_{i}\left(\vec{r},t\right)=\int f_{i}d\vec{v}_{i}$ and $\vec{u}_{i}\left(\vec{r},t\right)=\frac{1}{n_{i}}\int\vec{v}_{i}f_{i}d\vec{v}_{i}$
the density and the hydrodynamic velocity for the species $i$ respectively.
Here we have $\left\{ n_{1},...,n_{M},\vec{u},e\right\} $ as local
thermodynamic state variables. The distribution function per particle
$f_{i}$ for the constituent $i$ is the solution of the Boltzmann
equation for the species $i$. In the case of a mixture with $M$
components we have a set of $M$ Boltzmann equations, one per species,
and all of them coupled by the collisional terms. The details of corresponding
formalism can be found in Chapter 6 of Ref \cite{Ferziger}.

In the case of mixtures, the chaotic velocity plays crucial role,
not only because of the arguments discussed above, but also because
it helps to establish the diffusion of one species into the other.
The dissipative mass flux for the species $i$ is defined by
\begin{equation}
\vec{J}_{i}\left(\vec{r},t\right)=\frac{1}{n_{i}}\int\vec{c}_{i}fd\vec{c}_{i},\label{eq:2.13}
\end{equation}
and the relation
\begin{equation}
\sum_{i}\vec{J}_{i}=\vec{0},\label{eq:2.14}
\end{equation}
is a consequence of the definition (\ref{eq:2.11}). This last argument
is what actually gives sense to the idea of diffusion in mixtures
from the kinetic theory point of view. In the next section we will
show how these ideas are incorporated into relativistic kinetic theory
while preserving the covariance of the theory.

\section{The chaotic velocity in special relativistic kinetic theory}

Relativistic kinetic theory has been under development since 1911,
when the first proposal for an equilibrium distribution function was
presented \cite{JUTTNER}. The theory underwent a slow but steady
development and many relevant sources can be cited. The reader
is referred to the classical textbooks in Refs. \cite{Groot-Leewen-Weert,CERCIGNANI}
which abridge most of the work done in this direction. However, reviewing
the literature one can conclude that the concept of chaotic velocity
has been ignored up to this century. As mentioned above, this absence
was first pointed out in Ref. \cite{Sandoval physica A 2000} where
Lorentz transformations where firstly introduced, a subsequent work
rounded the idea \cite{Garcia Perciante 2012} which is currently
being applied with success in different frameworks (see for example
Refs. \cite{Garcia-Perciante Mendez Heat conduction ... revisited,tolman,benedicks,Val-3}).
In this section we show how the chaotic velocity is introduced in
special relativistic kinetic theory by considering Lorentz transformations
as a direct generalization of the Galilean ones used in the previous
section both for the single component gas as well as for the mixture.
Some details of this formalism can be found in Ref. \cite{Moratto Thesis}.

Following a procedure analogous to the one described in the previous
section, one starts form the relativistic Boltzmann equation, given
by
\begin{equation}
v^{\alpha}f_{,\alpha}=J\left(ff'\right).\label{eq:a}
\end{equation}
Multiplying Eq. (\ref{eq:a}) by the collisional invariants and integrating
over velocity space one obtains conservation equations for two fluxes,
the particle flux and the energy momentum tensor are given by
\begin{equation}
N^{\mu}=\int v^{\mu}fdv^{*},\label{b}
\end{equation}
and
\begin{equation}
T^{\mu\nu}=\int v^{\mu}v^{\nu}fdv^{*},\label{c}
\end{equation}
respectively. Here $dv^{*}$ is the invariant volume element \cite{CERCIGNANI}.
At this point, the introduction of the chaotic velocity is desirable
in order to distinguish, in a similar fashion as in the non-relativistic
case, the different components in each of these tensors. The corresponding
calculation together with a thorough discussion can be found in Ref.
\cite{Garcia Perciante 2012}. Here we discuss in depth the transformation
and the definition of the relevant variable as well as its invariance.
Let $U^{\mu}$ be the hydrodynamic four-velocity of the gas defined
by
\begin{equation}
U^{\mu}=\frac{dx^{\mu}}{d\tau},\label{eq:3.1}
\end{equation}
where $x^{\mu}$ is the four-position of the volume element of the
gas, as measured in a laboratory frame, and $\tau$ is the proper
time. Thus, $U^{\mu}$ is the four velocity of the fluid with respect
the laboratory inertial frame. The metric in this context is given
by
\begin{equation}
ds^{2}=dx^{2}+dy^{2}+dz^{2}-c^{2}dt^{2},\label{eq:3.2}
\end{equation}
in correspondence with Minkowski's metric. The components of $U^{\mu}$
are thus
\begin{equation}
U^{\mu}=\gamma_{u}\left(\vec{u},c\right),\label{eq:3.3}
\end{equation}
where $\gamma_{u}$ is the Lorentz factor, given by
\begin{equation}
\gamma_{u}=\frac{1}{\sqrt{1-\frac{u^{2}}{c^{2}}}},\label{eq:3.4}
\end{equation}
with $u^{2}=\vec{u}\cdot\vec{u}$. We can also define another four-velocity,
namely the four-velocity for one particle as observed in the laboratory
frame
\begin{equation}
v^{\mu}=\gamma_{v}\left(\vec{v},c\right),\label{eq:3.5}
\end{equation}
where clearly $\gamma_{v}=\left(1-v^{2}/c^{2}\right)^{-1/2}$ with
$v^{2}=\vec{v}\cdot\vec{v}$. It is also important to underline that both
four-velocities, $U^{\mu}$ and $v^{\mu}$ are time-like vectors,
i.e.
\begin{eqnarray}
U^{\mu}U_{\mu} & < & 0,\label{eq:3.7}\\
v^{\mu}v_{\mu} & < & 0,
\end{eqnarray}
and it is always possible to find a frame of reference in which the
world line of any of these vectors has only a temporal component \cite{Moller}.
For example, for the vector $U^{\mu}$ it is always possible to find
a frame of reference in which
\begin{equation}
U^{\mu}=\left[\vec{0},c\right].\label{eq:3.8}
\end{equation}
Then, for every inertial frame the invariant constructed with the
scalar contraction of the two four-velocities in question $U^{\mu}$
and $v^{\mu}$ reads
\begin{equation}
U^{\alpha}v_{\alpha}=\gamma_{u}\gamma_{v}\vec{u}\cdot\vec{v}-\gamma_{u}\gamma_{v}c^{2}.\label{eq:3.9}
\end{equation}
In particular, if we choose the frame defined by equation (\ref{eq:3.8}),
the invariant (\ref{eq:3.9}) reads
\begin{equation}
U^{\alpha}v_{\alpha}=-\gamma_{k}c^{2},\label{eq:3.10}
\end{equation}
where $\gamma_{k}$ is the Lorentz factor evaluated with the velocity
of the particle in such frame, whose magnitude we denote by $k$.
This quantity is, by its definition given in Eq. (\ref{eq:3.10}),
an invariant. Notice that this variable is also proportional to the
particles' kinetic energy measured in the comoving frame. Because
of this, it will be present in the integrals for various thermodynamic
quantities and it is thus worthwhile to point out that once evaluated
in the comoving frame, this factor is a Lorentz invariant in a similar
way as the rest mass and proper time are.

The frame defined by Eq. (\ref{eq:3.8}) is of great conceptual importance.
In this framework it is identified as the comoving frame, in a similar
way as was conceived in the non-relativistic case. Then, by following
the discussion of the last section, through the consideration of variables
and fluxes in this frame is how one can isolate the dissipative effects
from the total quantities which also include the convective ones.

\begin{figure}
\includegraphics[scale=0.5]{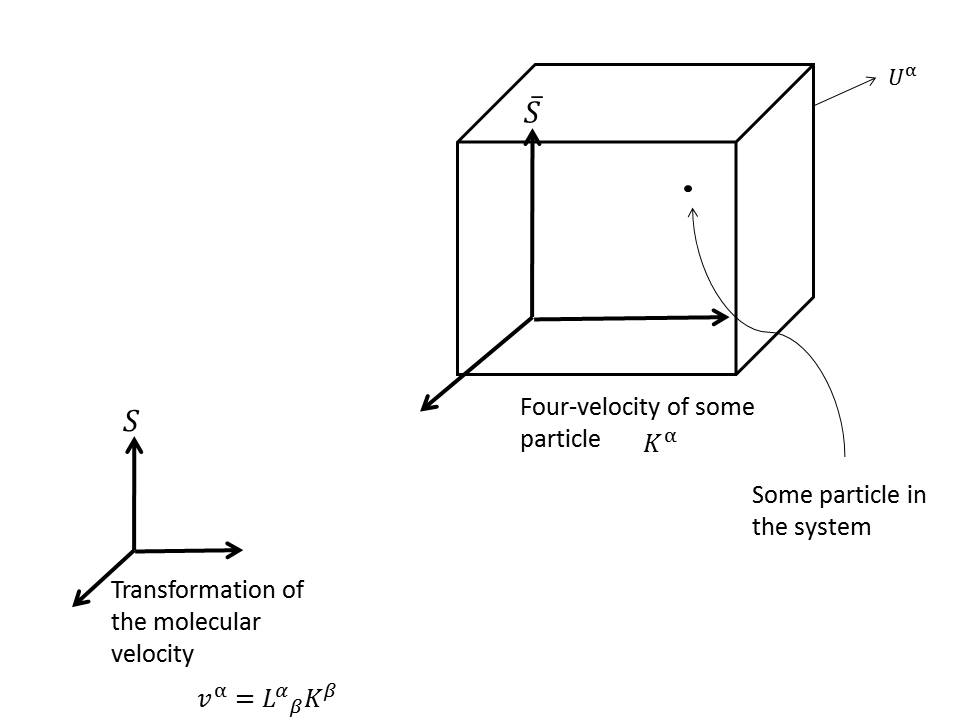}

\caption{
Comoving frame $\bar{S}$ moving with four-velocity $U^{\mu}$ respect
to the laboratory frame $S$. The chaotic four-velocity is $K^{\alpha}$.
}
\end{figure}

Figure 2 shows this frame in relation with the laboratory frame where
\begin{equation}
L_{\text{  }\beta}^{\alpha}=\frac{\partial \overline{x}^{\alpha}}{\partial x^{\beta}}\label{eq:3.11}
\end{equation}
is a Lorentz boost to the frame $\bar{S}$. Also, we have introduced
the particle four-velocity, as measured in the comoving frame $K^{\alpha}$,
in order to have a distinct notation for the chaotic velocity as in
the non-relativistic formalism. The relation between $K^{\alpha}$
and $v^{\alpha}$ is clearly given by
\begin{equation}
v^{\alpha}=L_{\text{  }\beta}^{\alpha}K^{\beta}.\label{eq:3.12}
\end{equation}
Equation (\ref{eq:3.12}) plays the role of Eq. (\ref{eq:2.6-1}).
With these definitions, one is able to write state variables and thermodynamic
fluxes as averages over chaotic quantities. For example for the internal
energy and heat flux one has
\begin{equation}
ne=mc^{2}\int\gamma_{k}^{2}fdK^{*},\label{d}
\end{equation}
and
\begin{equation}
q^{\beta}=mc^{2}L_{\beta}^{\mu}\int\gamma_{k}K^{\beta}fdK^{*}.\label{e}
\end{equation}
It is important to point out that these definitions are similar in
structure as the ones in the non-relativistic case and highlight the
physical nature of the corresponding quantities: the internal energy
as the average of the chaotic kinetic energy and the heat flux as
the flux of such quantity, measured in the comoving frame where mechanical
effects are not present. Notice that the heat flux definition includes
a Lorentz transformation. The deduction of these formulas can be found
to detail in Ref. \cite{Garcia Perciante 2012}. What we want to
point out here is the covariance of both quantities. Indeed, the internal
energy is an invariant, it does not depend on the observer and is
thus calculated in the comoving frame where it is the average of the
kinetic energy of the particles. On the other hand, the heat flux
is a tensor and thus transforms, in this case in a contravariant way.
Indeed, Eq. (\ref{e}) can be rewritten as $q^{\nu}=L_{\mu}^{\nu}Q^{\nu}$
where $Q^{\nu}$ is the heat flux as defined by Clausius
\cite{Brush}.

For the case of mixtures, the concept of chaotic velocity allows,
as in the previous section, for a clear definition of the diffusive
fluxes. As before, the comoving frame is replaced by the frame moving
with the barycentric velocity and chaotic velocities for each species
can be found by considering a Lorentz boost from the laboratory frame
to it. The fact that the species have different averages for chaotic
velocities gives rise, as before, to what we understand as diffusive
fluxes. To clarify this idea we consider some definitions from the
relativistic kinetic theory for mixtures \cite{Val-3,Val-4}. The
particle four-flow is defined by
\begin{equation}
N_{(i)}^{\alpha}\equiv m_{(i)}\int v_{(i)}^{\alpha}f_{(i)}dv_{(i)}^{*}\label{eq:3.12.1}
\end{equation}
for the $(i)$-th component of the gas, $m_{(i)}$ is the rest mass
of some particle of the species $(i)$ and $f_{(i)}$ the corresponding
solution of the Boltzmann equation. The total particle flow, that
is, the flow with the contribution of the $M$ species will be
\begin{equation}
N^{\alpha}=\sum_{(i)=1}^{M}N_{(i)}^{\alpha}.\label{eq:3.12.2}
\end{equation}
We can introduce the idea of chaotic velocity for the mixture in the
following way
\begin{equation}
v_{(i)}^{\alpha}=L_{\beta}^{\alpha}K_{(i)}^{\beta},\label{eq:3.12.3}
\end{equation}
where the Lorentz transformation $L_{\beta}^{\alpha}$ is constructed
with the barycentric velocity $U^{\alpha}$ defined by
\begin{equation}
nU^{\alpha}=\sum_{(i)=1}^{M}n_{(i)}U_{(i)}^{\alpha}\label{eq:3.12.4}
\end{equation}
with
\begin{equation}
n_{(i)}U_{(i)}^{\alpha}=\int v_{(i)}^{\alpha}f_{(i)}dv_{(i)}^{*}.\label{eq:3.12.5}
\end{equation}
By introducing Eq. (\ref{eq:3.12.3}) in Eq. (\ref{eq:3.12.1}), we
obtain a flow of particles evaluated in the comoving frame, which
is the diffusion of one species into the fluid, since it does not
involve convective effects. Thus, we have for the diffusive flux

\begin{equation}
J_{(i)}^{\mu}=m_{(i)}\int K_{(i)}^{\mu}f_{(i)}dK_{(i)}^{*}.\label{eq:3.12.6}
\end{equation}
Equation (\ref{eq:3.12.6}) satisfies the relation
\begin{equation}
\sum_{(i)=1}^{M}J_{(i)}^{\mu}=\left(\vec{0},n\right),\label{eq:3.12.7}
\end{equation}
where
\begin{equation}
n=\sum_{(i)}^{M}n_{(i)}.\label{eq:3.12.8}
\end{equation}
We can see from Eqs. (\ref{eq:3.12.6}) and (\ref{eq:3.12.7}) that
the diffusion has the same properties and physical meaning than in
the non-relativistic case. This argument reinforces the importance
of the use of chaotic velocity, Eq (\ref{eq:3.12.3}). These elements
are valid only in special relativity, some ideas regarding the general
relativistic generalization are explored in the next section.
\\

\section{The chaotic velocity in general coordinates}

In this section, the kinetic energy introduced before will be expressed
as an invariant representing the chaotic part of the particles' energy
for a general metric tensor, allowing for an extension to curved space-times.
In order to accomplish this task, we recall a theorem given by G. Ellis
which proposes one very particular relation between four-velocities
in order to introduce a relative velocity in a covariant fashion \cite{ellis articulo,ellis libro}.
Indeed, it is straightforward to show that given two time-like four-vectors $\left\{ v^{\mu},U^{\mu}\right\}$,
it is possible to find one space-like four-vector $S^{\mu}$ such
that
\begin{equation}
v^{\mu}=\eta\left(U^{\mu}+S^{\mu}\right)\label{eq:3.16}
\end{equation}
with
\begin{equation}
S^{\mu}U_{\mu}=0,\label{eq:3.17}
\end{equation}
and
\begin{equation}
\eta=\frac{1}{\sqrt{1-\frac{S_{\nu}S^{\nu}}{c^{2}}}}.\label{eq:3.18}
\end{equation}
For the sake of clarity, we recall that while a time-like four-vector
has the properties described in the previous section, a space-like
vector $S^{\mu}$ satisfies

\begin{equation}
S_{\nu}S^{\nu}>0,\label{eq:3.14}
\end{equation}
and it is always possible to find a reference
frame in which the temporal component of $S^{\mu}$ is zero, that
is
\begin{equation}
S^{\mu}=\left[\vec{s},0\right].\label{eq:3.15}
\end{equation}

This theorem is valid in general but in this case we associate $v^{\mu}$
with the molecular four-velocity and $U^{\mu}$ with the barycentric
or hydrodynamic four-velocity, both being time-like since $v^{\mu}v_{\mu}=U^{\mu}U_{\mu}=-c^{2}$.

Equation (\ref{eq:3.16}) is valid in any frame and, since $U^{\mu}$
is time-like, it is possible to evaluate it where $U^{\mu}=\left[\vec{0},c\right]$
which is precisely the fluid's comoving frame.
\begin{equation}
v_{\text{CM}}^{\alpha}=\eta\left(\left[\vec{0},c\right]+S_{\text{CM}}^{\alpha}\right),\label{eq:3.19}
\end{equation}
Now, in the special relativistic case addressed in the previous section,
the molecular velocity in the comoving frame is $K^{\alpha}$. Also,
since $S^{\mu}$ is orthogonal to $U^{\mu}$ (Eq. (\ref{eq:3.17})),
in the comoving frame it only has temporal components such that we
have
\begin{equation}
K^{\alpha}=\eta\left(\left[\vec{0},c\right]+\left[\vec{s},0\right]\right),\label{eq:3.19.1}
\end{equation}
which, since $K^{\alpha}=\gamma_{k}\left[\vec{k},c\right]$ yields
$\gamma_{k}=\eta$ and $\vec{s}=\vec{k}$. One concludes that in special relativity the $\gamma_{k}$ factor, which is proportional to the chaotic energy
and is thus required as a variable in order to express thermodynamic
quantities, can be written in general as
\begin{equation}
\gamma_{k}=\frac{1}{\sqrt{1-\frac{S_{\nu}S^{\nu}}{c^{2}}}},\label{eq:3.19.2}
\end{equation}
where $S^{\mu}$ is given by Eq. (\ref{eq:3.16}) and is a space-like
vector which in the comoving frame of the fluid, only has spatial
components which coincide with the chaotic velocity ones. This reinforces
the covariance of the calculations that have been worked in the literature
\cite{Garcia Perciante 2012,Garcia-Perciante Mendez Heat conduction ... revisited,Val-3,Val-4,Val-2}
in the framework of special-relativistic kinetic theory. Also, and
most importantly, this reasoning sheds light on a possible way to
extend these ideas to the general relativistic case.

For a general metric tensor $g_{\mu\nu}$, the kinetic energy of the
molecules measured in the comoving frame is given by $-mU^{\mu}v_{\mu}$.
Introducing the general decomposition given in Eq. (\ref{eq:3.16})
one obtains
\begin{equation}
mU^{\mu}v_{\mu}=-m\gamma c^{2},\label{eq:3.19.3}
\end{equation}
which leads to the interpretation of $\gamma=\eta$, given by Eq. (\ref{eq:3.18}),
as the generalization of $\gamma_{k}$ for a general metric. This
is a promising result since integrals representing state variables
and thermodynamic fluxes may be expressed in terms of this decomposition
having, in particular, the invariant $\gamma$ for energy quantities
available.
\\

\section{Conclusions}

In this paper we have revisited the concept of chaotic velocity in
the framework of relativistic kinetic theory by firstly recalling
some basic aspects of the non-relativistic case where such idea was
included since its early developments. Then, we addressed the special
relativistic case, emphasizing both its importance in the mixture
case as well as the covariance of the theory based on the invariance
of the relevant variable $\gamma_{k}$, representing the energy of
the particles in the comoving frame. In such a frame, in which the
hydrodynamic or barycentric velocity vanishes, the dissipative effects
are isolated since mechanical effects are not present.
\\

The importance of the chaotic velocity in the mixture case resides
in the fact that the state variable in such a case is the barycentric
velocity (not the hydrodynamic velocity for each species) and thus,
the diffusive fluxes of the species are relative to a frame comoving
with it. These fluxes are identified in this framework as the average
of the momentum of the particles measured in this comoving frame and
the sum of them vanishes in it.
\\

Regarding the covariance of the formalism, we provided with two solid
arguments. Firstly, what is a decomposition of velocities in the non-relativistic
case, is viewed as a reference frame transformation for the special
relativistic case. Such transformation consists in a Lorentz boost
which preserves the covariance. The relevant variable for the calculation
of fluxes and variables turns out to be proportional to the chaotic
component of the particles' energy which is expressed through an already
evaluated Lorentz factor $\gamma$. The second argument relies on
Ellis' theorem which introduces a space-like relative velocity. We
showed how $\gamma_{k}$ can be expressed in terms of the magnitude
of such a tensor. This idea yields an invariant expression for this
important quantity which can be generalized for a general metric.
\\

We conclude that the chaotic velocity is a valuable concept and can
be introduced in relativistic kinetic theory in a covariant fashion
which allows its formulation in a clear way by yielding the separation
of thermal and mechanical effects from the microscopic point of view.
The fact that the chaotic velocity can be defined as a tensor and
the corresponding $\gamma$ factor as an invariant for a general, not
necessarily flat, metric allows to foresee that the extension of this
conclusion for the general relativistic case is feasible. This idea
will be developed further in the near future.
\\

\section*{Acknowledgments}

The authors greatly appreciate the valuable comments from R. Sussman, A.
Sandoval-Villalbazo and G. Chac\'on-Acosta as well as fruitful discussions.
This work was supported by CONACyT through grant number CB2011/167563.



\bibliographystyle{aipproc}   


\end{document}